\documentclass[preprint,flushrt,eqsecnum,10pt]{aastex}
\usepackage{color}
\usepackage{url}
\newcommand{\kk}{\mathbf{k}}
\newcommand{\pp}{\mathbf{p}}

\newcommand{\qq}{\mathbf{q}}

\newcommand{\vv}{\mathbf{v}}

\begin{document}

\title{RELATION BETWEEN STANDARD PERTURBATION THEORY \\ AND REGULARIZED MULTI-POINT PROPAGATOR METHOD}
\author{Naonori S. Sugiyama\altaffilmark{1}}
\affil{Astronomical Institute, Graduate School of Science, Tohoku University, \\ 
        6-3, Aramakijiaoba, Sendai 980-8578, Japan}  
\email{sugiyama@astr.tohoku.ac.jp}
\altaffiltext{1}{Also at the Department of Astrophysical Sciences, Peyton Hall, Princeton University, Princeton, NJ 08544-1001, USA}
\and
\author{Toshifumi Futamase}
\affil{ Astronomical Institute, Graduate School of Science, Tohoku University} 
\affil{6-3, Aramakijiaoba, Sendai 980-8578, Japan}

\begin{abstract}

We investigate the relation between the regularized multi-propagator method, called ``Reg PT'', and the standard perturbation theory.
Reg PT is one of the most successful models to describe nonlinear evolution of dark matter fluctuations.
However, Reg PT is a mathematically unproven interpolation formula between the large-scale solution calculated by 
the standard perturbation theory and
the limiting solution in the small scale calculated by the multi-point propagator method.
In this paper, we give an alternative explanation for Reg PT in the context of the standard perturbation theory,
showing that Reg PT does not ever have more effective information on nonlinear matter evolution than the standard perturbation theory.
In other words, the solutions of the standard perturbation theory 
reproduce the results of $N$-body simulations better than those of Reg PT, especially at the high-$k$ region.
This fact means that the standard perturbation theory at the two-loop level 
is still one of the best predictions of the nonlinear power spectrum to date.
Nevertheless, the standard perturbation theory has not been preferred because of the divergent behavior of the solution at small scales.
To solve this problem, we also propose a modified standard perturbation theory which avoids the divergence.

\end{abstract}


\maketitle
\section{Introduction}

What is the best model to predict the nonlinear matter power spectrum ?
In this paper, we show that the standard perturbation (SPT) theory with two-loop corrections 
is still one of the best models, even though various modified perturbation theories have been proposed in the last 5-10 yr.
In doing so, we give the relation between the SPT and the regularized multi-propagator method.

One of the key quantities in modern cosmology is the matter power spectrum because it contains 
a lot of important information on evolution of the universe as well as on structure formation. 
Recent observation of the baryon acoustic oscillations (BAO) in the power spectrum 
provides a new method to precisely restrict cosmological parameters~\citep{Eisenstein/etal:2005}.

Recent progress on cosmological observation greatly motivates 
various theoretical studies of accurate calculation of the nonlinear matter power spectrum.
At present, one of the most successful models is given by the multi-point propagator method 
($\Gamma$-expansion method) with the regularized treatment of propagators
presented by \citet{Bernardeau:2008fa,Bernardeau:2011dp}, called ``Reg PT'' in~\citet{Taruya:2012ut}.
In particular, \citet{Taruya:2012ut} computed the solution of Reg PT with two-loop corrections,
giving theoretical predictions of the power spectrum which agree well
with $N$-body simulations at BAO scales.

Various modified perturbation theories partially sum up high-order terms in the SPT,
called ``resummation theories''.
Since SPT has exact but formal solutions at any order in perturbation theory,
any resummation theory should be represented in the context of SPT.
The coefficients of the $\Gamma$-expansion can easily be represented by the kernel function in SPT.
This is why we consider the relation between SPT and the $\Gamma$-expansion method.

We also proposed a method called the ``Wiener Hermite (WH) expansion``
which gives an accurate power spectrum over relatively wide range of wavenumber~\citep{Sugiyama:2012pc}.
There, we established the relation between the WH expansion and SPT, and showed the equivalence between 
the WH expansion and the $\Gamma$-expansion method~\citep{Bernardeau:2008fa,Bernardeau:2011dp,Taruya:2012ut}.
Furthermore, by using an approximation of the kernel functions in SPT,
we derived the exponentially damping behavior of the power spectrum at each order of the WH expansion ($\Gamma$-expansion) 
which has been known in the renormalized perturbation theory (RPT; \citet{Crocce:2005xz,Crocce:2005xy,Crocce:2007dt}).

We extend the approximation method of the kernel functions introduced in our previous work~\citep{Sugiyama:2012pc}.
As a result, we give an alternative explanation for Reg PT, and propose a natural extension of Reg PT.
The extended Reg PT gives the SPT solution with negligible correction terms.
Thus, we find that the Reg PT solution does not ever have more effective information on the evolution of dark matter than the SPT solution.
In addition, we also propose a modified version of the SPT solution which avoids the divergent behavior of the SPT solution 
at small scales.

The $N$-body simulation results used in this paper were presented by \citet{Valageas:2010yw}.
These results and initial conditions at $z_{\rm ini} = 99$ 
were created by the public $N$-body codes {\it GADGET2} and {\it 2LPT} code, respectively~\citep{Springel:2005mi,Crocce:2006ve}.
These $N$-body simulations contain $2048^3$ particles 
and were computed by combining the results with different box sizes $2048 h^{-1}$ ${\rm Mpc}$ and $4096$ $h^{-1}{\rm Mpc}$, called 
$L11$-$N11$ and $L12$-$N11$. 
The cosmological parameters we used were presented by 
the {\it Wilkinson Microwave Anisotropy Probe} five year release~(\citet{Komatsu:2008hk},
$\Omega_m = 0.279$, $\Omega_{\Lambda} = 0.721$, $\Omega_b = 0.046$, $h = 0.701$, $n_s = 0.96$ and $\sigma_8=0.817$).
We used the program which is available on Taruya's homepage to compute the predicted power spectra
\footnote{ \url{ http://www-utap.phys.s.u-tokyo.ac.jp/~ataruya/}}.

This paper is organized as follows.
In Section~\ref{Review}, the relation between SPT and the $\Gamma$-expansion is reviewed.
Section~\ref{Ap_Kernel} proves the approximation of the kernel functions.
Using the approximated kernel functions, the alternative explanation for Reg PT is given in Section~\ref{Proof}.
In Section~\ref{Extension_of_RegPT}, it is proposed that the natural extension of Reg PT 
is equivalent to the solution of SPT with some correction terms.
We also present the modified version of the SPT solution in Section~\ref{Ex_SPT}.
There, we reconfirm that the two-loop SPT solutions are better than those of the two-loop Reg PT and the closure theory,
and show that the divergent behavior of the solution in SPT is indeed removed by our modification.
In Section~\ref{Conclusion}, we summarize and discuss our results.

\section{Review of SPT and the $\Gamma$-expansion method}
\label{Review}

We shall briefly review SPT. 
In SPT, we choose a cosmological model with $f=\Omega_m^{1/2}$ 
where $f \equiv d \ln D/ d \ln a $ is the linear growth rate with $D$ and $a$ being the linear growth factor and the scale factor.
As shown by \citet{1991MNRAS.251..128L},
this model is a good approximation in practice.
In this situation, the density perturbation and the velocity divergence of dark matter are expanded as~\citep{Bernardeau:2001qr},
\begin{equation}
		\delta(z,\kk) = \sum_{n=1}^{\infty} D^{n} \delta_{n}(\kk), \qquad 
		\theta(z,\kk) = -aHf\sum_{n=1}^{\infty} D^n \theta_n(\kk),
\end{equation}
where $H$ is the Hubble parameter, and $\delta_n$ and $\theta_n$ are time-independent quantities with a Fourier mode $\kk$.
The velocity divergence is defined as $\theta \equiv \nabla \cdot \vv$, where $\vv$ is the velocity of dark matter.
The $n$th-order solutions in SPT are given by,
\begin{eqnarray}
		\delta_n(\kk) &=&  \int \frac{d^3p_1}{(2\pi)^3} \cdots \int \frac{d^3p_n}{(2\pi)^3} 
		(2\pi)^3 \delta_{\rm D}(\kk-\pp_{[1,n]}) F_{n}([\pp_1,\pp_n]) \delta_{\rm L}(\pp_1)\cdots \delta_{\rm L}(\pp_n), \nonumber \\
		\theta_n(\kk) &=&  \int \frac{d^3p_1}{(2\pi)^3} \cdots \int \frac{d^3p_n}{(2\pi)^3} 
		(2\pi)^3 \delta_{\rm D}(\kk-\pp_{[1,n]}) G_{n}([\pp_1,\pp_n]) \delta_{\rm L}(\pp_1)\cdots \delta_{\rm L}(\pp_n), 
\label{sol_FG}
\end{eqnarray}
where $\pp_{[1,n]}\equiv\pp_1+\cdots+\pp_n$, $F_n(\pp_1,\dots,\pp_n) \equiv F_n([\pp_1,\pp_n])$, and 
$\delta_{\rm D}$ is the Dirac delta function.
Linearized quantities, such as $\delta_{\rm L}$ and $\theta_{\rm L}$, are denoted by the subscript $L$.
The kernel functions $F$ and $G$ are constructed from the mode coupling functions 
$\alpha(\kk_1,\kk_2) \equiv (\kk_1+\kk_2)\cdot\kk_2/k_1^2$ and $\beta(\kk_1,\kk_2)\equiv |\kk_1+\kk_2|^2 (\kk_1\cdot\kk_2)/2k_1^2k_2^2$
according to the following recursion relation:
\begin{eqnarray}
		 n F_{n}([\pp_1,\pp_{n}]) - G_{n}([\pp_1,\pp_{n}]) 
		&=& \sum_{m=1}^{n-1}G_m([\qq_1,\qq_m])\alpha(\qq_{[1,m]},\qq_{[m+1,n]})F_{n-m}([\qq_{m+1},\qq_{n}]), \nonumber \\
		-\frac{3}{2}F_{n}([\pp_1,\pp_{n}]) +\frac{2n+1}{2}G_{n}([\pp_1,\pp_{n}]) 
		&=& \sum_{m=1}^{n-1}G_m([\qq_1,\qq_m])\beta(\qq_{[1,m]},\qq_{[m+1,n]})G_{n-m}([\qq_{m+1},\qq_{n}]), \nonumber \\
		\label{FG}
\end{eqnarray}
where $F_1 = G_1 = 1$, and 
the wavenumber vectors $\{\qq_1,\dots,\qq_n\} = \{\pp_1,\dots,\pp_n\}$ in the right hand side are completely-symmetrized.

The relation between the $r$th-order coefficient in the $\Gamma$-expansion and the kernel function is given by
\citep{Bernardeau:2008fa,Sugiyama:2012pc}
\begin{eqnarray}
		\Gamma^{(r)}(z,[\kk_1,\kk_r])
		\equiv 	D^r\Gamma^{(r)}_{\rm tree}([\kk_1,\kk_r]) 
		+ \sum_{n=1}^{\infty} D^{r+2n} \Gamma_{\rm n\mathchar`-loop}^{(r)}([\kk_1,\kk_r]),
\end{eqnarray}
with
\begin{eqnarray}
		&&\Gamma_{\rm tree}^{(r)}([\kk_1,\kk_r])\equiv F_r([\kk_1,\kk_r]), \nonumber \\
		&&\Gamma^{(r)}_{\rm n\mathchar`-loop}([\kk_1,\kk_r]) \nonumber \\
		&&\equiv
       \frac{1}{r!}\frac{(r+2n)!}{2^n n!} 
          \int \frac{d^3p_1}{(2\pi)^3} \cdots  \int \frac{d^3p_n}{(2\pi)^3}
		F_{r+2n}([\kk_1,\kk_r],\pp_1,-\pp_1,\dots,\pp_n,-\pp_n)P_{\rm L}(p_1)\cdots P_{\rm L}(p_n), \nonumber \\
		\label{G-r}
\end{eqnarray}
where $P_{\rm L}$ is the present linear power spectrum.
The $n$-loop correction to the $r$th-order coefficient of the $\Gamma$-expansion
$\Gamma_{\rm n\mathchar`-loop}^{(r)}$ is of ${\cal O}( (P_{\rm L})^n)$,
and involves $n$-dimensional integrals.
The full nonlinear matter power spectrum is given by
\begin{eqnarray}
		P(z,k) = \sum_{r=1}^{\infty} P_{\rm \Gamma}^{(r)}(z,k),
\end{eqnarray}
where $P_{\rm \Gamma}$ is the $r$th-order contribution to the power spectrum in the $\Gamma$-expansion, defined as
\begin{equation}
		P_{\rm \Gamma}^{(r)}(z,k) \equiv  r! \int \frac{d^3k_1}{(2\pi)^3} \cdots  \int \frac{d^3k_r}{(2\pi)^3}
		(2\pi)^3 \delta_D(\kk-\kk_{[1,r]}) \left[\Gamma^{(r)}(z,[\kk_1,\kk_r])  \right]^2P_{\rm L}(k_1) \cdots P_{\rm L}(k_r).
	\label{power-r}
\end{equation}
Note that $P_{\rm \Gamma}$ which comprises $\Gamma_{\rm n\mathchar`-loop}^{(r)}\times \Gamma_{\rm m\mathchar`-loop}^{(r)}$ 
is of ${\cal O}\left( (P_{\rm L})^{r+n+m} \right)$,
and has $(r+n+m-1)$-dimensional integrals, also called the $(r+n+m-1)$-loop correction term.

The velocity divergence $\theta$ satisfies the same relation as the above by replacing $F$ with $G$.

\section{Approximation of Kernel Functions}
\label{Ap_Kernel}

Since the $\Gamma$-expansion method is rewritten by the kernel functions in SPT, 
the problem to solve the nonlinear evolution of dark matter eventually reduces to compute the kernel functions.
Therefore, we develop an approximation of the kernel functions.

We shall impose the condition that the amplitude of a Fourier mode $\pp$ in $F_n$ and $G_n$ is much smaller than that of the others,
\begin{equation}
		|\pp_i| \gg |\pp| \to 0, \quad {\rm for} \ \{ i = 1,\dots,n-1\}.
\end{equation}
In this condition, we prove the following approximation of the kernel functions:
\begin{eqnarray}
		F_n([\pp_1,\pp_{n-1}],\pp) &\to& \frac{1}{n}\left(\frac{\pp_{[1,n-1]}\cdot \pp}{p^2} \right) F_{n-1}([\pp_1,\pp_{n-1}]), \nonumber \\
		G_n([\pp_1,\pp_{n-1}],\pp) &\to& \frac{1}{n}\left(\frac{\pp_{[1,n-1]}\cdot \pp}{p^2} \right) G_{n-1}([\pp_1,\pp_{n-1}]).
		\label{ap_FG}
\end{eqnarray}
We provide proof of this by the recursion relation in Equation~(\ref{FG}) and induction in $n$.
Note that these expressions are slightly different from 
and more generalized than those presented in our previous work~\citep{Sugiyama:2012pc}.

First, the kernel functions $F_2$ and $G_2$ clearly satisfy Equation~(\ref{ap_FG}) for $|\pp_2| \gg |\pp_1| \to0$:
\begin{eqnarray}
		F_2(\pp_1,\pp_2)\big|_{p_1\to0}
		&\equiv& \frac{5}{7} + \frac{1}{2}\frac{\pp_1 \cdot \pp_2}{p_1p_2}\left( \frac{p_1}{p_2} + \frac{p_2}{p_1} \right)
		+ \frac{2}{7}\frac{(\pp_1\cdot\pp_2)^2}{p_1^2p_2^2}
		\to \frac{1}{2} \left( \frac{\pp_1\cdot\pp_2}{p_1^2} \right)F_1, \nonumber \\
        G_2(\pp_1,\pp_2)\big|_{p_1\to0}
		&\equiv& \frac{3}{7} + \frac{1}{2}\frac{\pp_1 \cdot \pp_2}{p_1p_2}\left( \frac{p_1}{p_2} + \frac{p_2}{p_1} \right)
		+ \frac{4}{7}\frac{(\pp_1\cdot\pp_2)^2}{p_1^2p_2^2}
		\to \frac{1}{2} \left( \frac{\pp_1\cdot\pp_2}{p_1^2} \right)G_1.
		\label{F_2G_2}
\end{eqnarray}

The next step is to prove the following relation for $|\pp|\to0$:
\begin{eqnarray}
		F_{n+1}([\pp_1,\pp_{n}],\pp) &\to& \frac{1}{n+1}\left(\frac{\pp_{[1,n]}\cdot \pp}{p^2} \right) F_{n}([\pp_1,\pp_{n}]), \nonumber \\
		G_{n+1}([\pp_1,\pp_{n}],\pp) &\to& \frac{1}{n+1}\left(\frac{\pp_{[1,n]}\cdot \pp}{p^2} \right) G_{n}([\pp_1,\pp_{n}]),
		\label{F_n+1G_n+1}
\end{eqnarray}
with the assumption that $F_i$ and $G_i$ for $\{i = 1,\dots,n\}$ satisfy Equation~(\ref{ap_FG}).
For this purpose, we rewrite the first line in Equation~(\ref{FG}) as follows:
\begin{eqnarray}
		&& (n+1) F_{n+1}([\pp_1,\pp_{n}],\pp) - G_{n+1}([\pp_1,\pp_{n}],\pp) \nonumber \\
		&=&\sum_{m=1}^{n-1}  \left( \frac{m+1}{n+1} \right)
		G_{m+1}([\qq_1,\qq_{m}],\pp)\alpha(\qq_{[1,m]}+\pp,\qq_{[m+1,n]})F_{n-m}([\qq_{m+1},\qq_{n}]) \nonumber \\
		&& + \sum_{m=1}^{n-1} \left( \frac{n+1-m}{n+1}  \right)
		G_m([\qq_1,\qq_{m}])\alpha(\qq_{[1,m]},\qq_{[m+1,n]}+\pp)F_{n+1-m}([\qq_{m+1},\qq_{n}],\pp) \nonumber \\
		&& + \frac{1}{n+1} \alpha(\pp,\pp_{[1,n]}) F_n([\pp_1,\pp_n])
		   +  \frac{1}{n+1}\alpha(\pp_{[1,n]},\pp) G_n([\pp_1,\pp_n]).
		\label{FG2}
\end{eqnarray}
This equation is approximated as
\begin{eqnarray}
		&& (n+1) F_{n+1}([\pp_1,\pp_{n}],\pp) - G_{n+1}([\pp_1,\pp_{n}],\pp) \nonumber \\
		& \to& \frac{1}{n+1} \left( \frac{\pp_{[1,n]}\cdot\pp}{p^2} \right)
		\sum_{m=1}^{n-1}G_{m}([\qq_1,\qq_{m}])\alpha(\qq_{[1,m]},\qq_{[m+1,n]})F_{n-m}([\qq_{m+1},\qq_{n}]) \nonumber \\
		&& + \frac{1}{n+1} \left( \frac{\pp_{[1,n]}\cdot\pp}{p^2} \right) F_n([\pp_1,\pp_n])
		   +  \frac{1}{n+1} G_n([\pp_1,\pp_n]),
		   \label{FG3}
\end{eqnarray}
where we used the following relations for $|\pp|\to0$:
\begin{eqnarray}
		G_{m+1}([\qq_1,\qq_m],\pp) &\to& \frac{1}{m+1} \left( \frac{\qq_{[1,m]}\cdot\pp}{p^2} \right)G_m([\qq_1,\qq_m]), \nonumber \\
		F_{n+1-m}([\qq_{m+1},\qq_n],\pp) &\to& \frac{1}{n+1-m} \left( \frac{\qq_{[m+1,n]}\cdot\pp}{p^2} \right),
		F_{n-m}([\qq_m,\qq_n]),\nonumber \\
	    \alpha(\qq_{[1,m]}+\pp,\qq_{[m+1,n]}) &\to&  \alpha(\qq_{[1,m]},\qq_{[m+1,n]}), \nonumber \\
        \alpha(\qq_{[1,m]},\qq_{[m+1,n]}+\pp) &\to&   \alpha(\qq_{[1,m]},\qq_{[m+1,n]}), \nonumber \\
		\alpha(\pp_{[1,n]},\pp) &\to& 1.
\end{eqnarray}
Using the first line in Equation~(\ref{FG}) and ignoring the last term in Equation~(\ref{FG3}),
we find
\begin{eqnarray}
		&& (n+1) F_{n+1}([\pp_1,\pp_{n}],\pp) - G_{n+1}([\pp_1,\pp_{n}],\pp)  \nonumber \\
		&&\to \frac{1}{n+1} \left( \frac{\pp_{[1,n]}\cdot\pp}{p^2} \right)
		     \left[ nF_n([\pp_1,\pp_n]) - G_n([\pp_1,\pp_n]) \right]
			 + \frac{1}{n+1} \left( \frac{\pp_{[1,n]}\cdot\pp}{p^2} \right) F_n([\pp_1,\pp_n]) \nonumber \\
			 && = \frac{1}{n+1}\left( \frac{\pp_{[1,n]}\cdot\pp}{p^2} \right)
		     \left[ (n+1) F_n([\pp_1,\pp_n]) - G_n([\pp_1,\pp_n])\right].
			 \label{ap_FG_delta}
\end{eqnarray}
For the second line in Equation~(\ref{FG}), similar calculations lead to
\begin{eqnarray}
		&& -\frac{3}{2}F_{n+1}([\pp_1,\pp_{n}],\pp) +\frac{2n+3}{2}G_{n+1}([\pp_1,\pp_{n}],\pp)  \nonumber \\
		&& \to  \frac{1}{n+1}\left( \frac{\pp_{[1,n]}\cdot\pp}{p^2} \right)
		\left[ -\frac{3}{2}F_{n}([\pp_1,\pp_{n}]) +\frac{2n+3}{2}G_{n}([\pp_1,\pp_{n}])  \right],
		\label{ap_FG_theta}
\end{eqnarray}
where we used 
\begin{equation}
		\beta(\pp_{[1,n]},\pp) \to \frac{\pp_{[1,n]}\cdot\pp}{2p^2}, \ \ \mbox{for $|\pp|\to0$}.
\end{equation}
Combining Equations~(\ref{ap_FG_delta}) and (\ref{ap_FG_theta}), we finally obtain Equation~(\ref{F_n+1G_n+1}).
This concludes the proof.

For $p_{m+1},\cdots,p_n\to0$, Equation~(\ref{ap_FG}) is revised as
\begin{eqnarray}
		F_n([\pp_1,\pp_{n}])\big|_{p_{m+1},\dots,p_n\to0} &\to& \frac{m!}{n!}\left(\frac{\pp_{[1,m]}\cdot \pp_{m+1}}{p^2_{m+1}} \right) 
     	   \cdots \left(\frac{\pp_{[1,m]}\cdot \pp_{n}}{p^2_{n}} \right) F_{m}([\pp_{1},\pp_{m}]), 
\end{eqnarray}
where $n > m$.
Thus, we find that $F_n$ is approximated using $F_{m}$ at a lower order $m$ than $n$.
The case of $m=1$ leads to the Zel'dovich approximation in the limit that the total momentum is $p_1$:
\begin{eqnarray}
		F_n(\pp_1,\pp_{n}])|_{p_2,\dots,p_n\to0} \to
           \frac{1}{n!}\left(\frac{\pp_{1}\cdot \pp_{2}}{p^2_{2}} \right) \cdots \left(\frac{\pp_{1}\cdot \pp_{n}}{p^2_{n}} \right) .
\end{eqnarray}
Therefore, we interpret the approximation of Equation~(\ref{ap_FG}) as an extension of the Zel'dovich approximation.

\section{Reg PT}
\label{Proof}
Now, we are ready to derive Reg PT \citep{Bernardeau:2008fa,Bernardeau:2011dp,Taruya:2012ut}.
We give an alternative explanation for Reg PT by the relation between SPT and the $\Gamma$-expansion [Equation~(\ref{G-r})]
and the approximation of the kernel functions [Equation~(\ref{ap_FG})].

Before we proceed, we show that the $r$th-order coefficient of the $\Gamma$-expansion at the $n$-loop level 
$\Gamma_{\rm n\mathchar`-loop}^{(r)}$ is approximately represented by $\Gamma_{\rm m\mathchar`-loop}^{(r)}$ at a lower $m$-loop level 
than the $n$-loop level: 
\begin{eqnarray}
		\Gamma_{\rm n\mathchar`-loop}^{(r)}([\kk_1,\kk_r])\big|_{p_{m+1},\cdots,p_n\to0}
		\to \frac{m!}{n!} \left( -\frac{k^2\sigma_v^2}{2} \right)^{n-m} \Gamma_{\rm m\mathchar`-loop}^{(r)}([\kk_1,\kk_r]),
		\label{G:n-mloop}
\end{eqnarray}
where $\sigma_v^2$ is the velocity dispersion of dark matter: $\sigma_v^2\equiv \int dp P_L(p)/6\pi $.
This is shown by substituting the following approximated kernel functions into Equation~(\ref{G-r}):
\begin{eqnarray}
		&&F_{r+2n}([\kk_1,\kk_r],\pp_1,-\pp_1,\cdots,\pp_n,-\pp_n)\big|_{p_{m+1},\cdots,p_n \to0} \nonumber \\
		&\to& \frac{(r+2m)!}{(r+2n)!} (-1)^{n-m} 
		\left( \frac{\kk_{[1,r]}\cdot \pp_{m+1}}{p_{m+1}} \right)^2 \cdots \left( \frac{\kk_{[1,r]}\cdot \pp_n}{p_n^2} \right)^2
		F_{r+2m}\left( [\kk_1,\kk_r],\pp_1,-\pp_1,\cdots,\pp_m,-\pp_m \right). \nonumber \\
\end{eqnarray}

To lead Reg PT at the one-loop level, we need the cases of $m=0$ and $m=1$ in Equation~(\ref{G:n-mloop}):
\begin{eqnarray}
		\Gamma_{\rm n\mathchar`-loop}^{(r)}([\kk_1,\kk_r])\big|_{p_2,\cdots,p_n\to0}
		&=& \frac{1}{n!}\left( -\frac{k^2\sigma_v^2}{2} \right)^{n-1} \Gamma_{\rm 1\mathchar`-loop}^{(r)}([\kk_1,\kk_r]), \nonumber \\
		\Gamma_{\rm n\mathchar`-loop}^{(r)}([\kk_1,\kk_r])\big|_{p_1,\cdots,p_n\to0} 
		&=& \frac{1}{n!}\left( -\frac{k^2\sigma_v^2}{2} \right)^{n} \Gamma_{\rm tree}^{(r)}([\kk_1,\kk_r]).
		\label{eq:1}
\end{eqnarray}
From Equation~(\ref{eq:1}), we derive an approximated expression of $\Gamma_{\rm n-loop}^{(r)}$ as follows:
\begin{eqnarray}
		\Gamma_{\rm n\mathchar`-loop}^{(r)}([\kk_1,\kk_r]) &\to& n\Gamma_{\rm n\mathchar`-loop}^{(r)}([\kk_1,\kk_r])\big|_{p_2,\dots,p_n \to0}
		 - (n-1) \Gamma_{\rm n\mathchar`-loop}^{(r)}([\kk_1,\kk_r])\big|_{p_1,\cdots,p_n \to0} \nonumber \\
		 &=& \frac{n}{n!}\left( -\frac{k^2 \sigma_v^2}{2} \right)^{n-1}\Gamma_{\rm 1\mathchar`-loop}^{(r)}([\kk_1,\kk_r])
		- \frac{n-1}{n!}\left( -\frac{k^2 \sigma_v^2}{2} \right)^{n}\Gamma_{\rm tree}^{(r)}([\kk_1,\kk_r]), \nonumber \\
		\label{del_r+2n}
\end{eqnarray}
where we multiply the first term in the right hand side by $n$ because 
we choose a Fourier mode $\pp_1$ from $n$ Fourier modes $\pp_i$ $\{i=1,\cdots,n\}$.
Furthermore, we need to subtract $(n-1)\Gamma_{\rm n-loop}^{(r)}|_{p_1,\cdots,p_n \to0}$
because the first term $n\Gamma_{\rm n\mathchar`-loop}^{(r)}|_{p_2,\dots,p_n\to0}$ 
integrates the same region defined by $p_1,\dots,p_n\to0$ $n$ times in the multiple integral in Equation~(\ref{G-r}).
This expression is satisfied even in the case of $n=0$: $\Gamma_{\rm 0\mathchar`-loop}^{(r)} = \Gamma_{\rm tree}^{(r)}$.
Substituting Equation~(\ref{del_r+2n}) into Equation~(\ref{G-r}),
we find Reg PT with the one-loop corrections:
\begin{eqnarray}
		&&\Gamma^{(r)}(z,[\kk_1,\kk_r]) \nonumber \\
		&&\to \sum_{n=0}^{\infty} D^{r+2n} 
		\left[ \frac{n}{n!}\left( -\frac{k^2 \sigma_v^2}{2} \right)^{n-1}\Gamma_{\rm 1\mathchar`-loop}^{(r)}([\kk_1,\kk_r])
		- \frac{n-1}{n!}\left( -\frac{k^2 \sigma_v^2}{2} \right)^{n}\Gamma_{\rm tree}^{(r)}([\kk_1,\kk_r])\right]
		\nonumber \\
		&=& \exp\left(   -\frac{k^2 D^2\sigma_v^2}{2} \right)
		D^r\left[ \Gamma_{\rm tree}^{(r)}([\kk_1,\kk_r]) + \left(D^2\Gamma_{\rm 1\mathchar`-loop}^{(r)}[\kk_1,\kk_r]
		+ \frac{k^2 D^2 \sigma_v^2}{2} \Gamma_{\rm tree}^{(r)}([\kk_1,\kk_r])   \right) \right]. \nonumber \\
		\label{Reg1}
\end{eqnarray}

As in the case of Reg PT at the one-loop level, we are further able to show the following expression:
\begin{eqnarray}
		&&\Gamma_{\rm n\mathchar`-loop}^{(r)}([\kk_1,\kk_r])  \nonumber \\
		&\to&
		 \frac{1}{(n-2)!}\left( -\frac{k^2 \sigma_v^2}{2} \right)^{n-2}
		 \Bigg[\Gamma_{\rm 2\mathchar`-loop}^{(r)}([\kk_1,\kk_r]) + \left( \frac{k^2\sigma_v^2}{2} \right)
		 \Gamma_{\rm 1\mathchar`-loop}^{(r)}([\kk_1,\kk_r]) 
		 + \frac{1}{2}\left( \frac{k^2\sigma_v^2}{2} \right)^2 \Gamma_{\rm tree}^{(r)}([\kk_1,\kk_r]) \Bigg] \nonumber \\
		&& +  \frac{1}{(n-1)!}\left( -\frac{k^2 \sigma_v^2}{2} \right)^{n-1}
		\Bigg[\Gamma_{\rm 1\mathchar`-loop}^{(r)}([\kk_1,\kk_r]) 
		+ \left( \frac{k^2\sigma_v^2}{2} \right) \Gamma_{\rm tree}^{(r)}([\kk_1,\kk_r]) \Bigg]
		\nonumber \\
		&& +  \frac{1}{n!}\left( -\frac{k^2 \sigma_v^2}{2} \right)^{n} \Gamma_{\rm tree}^{(r)}([\kk_1,\kk_r]).
\end{eqnarray}
This results in Reg PT with the two-loop corrections:
\begin{eqnarray}
	&&	\Gamma^{(r)}(z,[\kk_1,\kk_r]) \nonumber \\
	&&	\to \exp\left(  -\frac{k^2 D^2\sigma_v^2}{2} \right)
		D^r\Bigg[ \Gamma_{\rm tree}^{(r)}([\kk_1,\kk_r]) + 
		\left(  D^2\Gamma_{\rm 1\mathchar`-loop}^{(r)}([\kk_1,\kk_r]) 
		+ \frac{k^2 D^2\sigma_v^2}{2}  \Gamma_{\rm tree}^{(r)}([\kk_1,\kk_r]) \right) 
		\nonumber \\
		&& + \left( D^4\Gamma_{\rm 2\mathchar`-loop}^{(r)}([\kk_1,\kk_r]) 
		+ \left( \frac{k^2 D^2\sigma_v^2}{2}  \right)\Gamma_{\rm 1\mathchar`-loop}^{(r)}([\kk_1,\kk_r])
		+ \frac{1}{2} \left(\frac{k^2 D^2\sigma_v^2}{2}  \right)^2 \Gamma_{\rm tree}^{(r)}([\kk_1,\kk_r])  \right)	  \Bigg]. \nonumber \\
	\label{Reg2}
\end{eqnarray}

\section{Extension of Reg PT}
\label{Extension_of_RegPT}

In this section, we propose a natural extended version of Reg PT.
In Section~\ref{Proof},
we only applied the approximated kernel functions to multiple integrals involved in the coefficients of the $\Gamma$-expansion,
such as the integrals of $\pp_1$, \dots, $\pp_n$ in Equation~(\ref{G-r}).
Here, we further use the approximated kernel functions to calculate the multiple integrals 
in the $r$th-order contribution to the power spectrum in the $\Gamma$-expansion,
such as the integrals of $\kk_1$, \dots, $\kk_r$ in Equation~(\ref{power-r}).
In other words, we apply our kernel approximation method to compute high-order terms of the $\Gamma$-expansion with Reg PT.

The authors in \cite{Taruya:2012ut} computed $P_{\rm \Gamma}^{(r)}$ up to the third order.
The terms at the first, second, and third order in the $\Gamma$-expansion 
were calculated using Reg PT at the two-loop, one-loop (Equations~(\ref{Reg2}) and (\ref{Reg1})), and tree level, respectively
\citep[see][Equations~(23)-(26)]{Taruya:2012ut}.
On the other hand, we evaluate any order of the $\Gamma$-expansion using either Equation~(\ref{Reg1}) or Equation~(\ref{Reg2})
to keep consistency of the approximation.

The important relation to extend Reg PT is 
that the $r$th-order coefficient in the $\Gamma$-expansion 
is approximated by one at a lower order $m$ than $r$:
\begin{eqnarray}
		\Gamma_{\rm n\mathchar`-loop}^{(r)}([\kk_1,\kk_r])\big|_{k_{m+1},\dots,k_r\to0} 
		\to \frac{m!}{r!}\left( \frac{\kk_{[1,m]}\cdot \kk_{m+1}}{k_{m+1}^2} \right)
		\cdots\left( \frac{\kk_{[1,m]}\cdot \kk_{m}}{k_{m}^2} \right) \Gamma_{\rm n\mathchar`-loop}^{(m)}([\kk_1,\kk_m]),
		\label{ap_G}
\end{eqnarray}
where we used the following approximated kernel function,
\begin{eqnarray}
		&&F_{r+2n}([\kk_1,\kk_r],\pp_1,-\pp_1,\dots,\pp_n,-\pp_n)\big|_{k_{m+1},\dots,k_n\to0} \nonumber \\
		&=&  \frac{(m+2n)!}{(r+2n)!}\left( \frac{\kk_{[1,m]}\cdot \kk_{m+1}}{k_{m+1}^2} \right)
		\cdots \cdots\left( \frac{\kk_{[1,m]}\cdot \kk_{m}}{k_{m}^2} \right)F_{m+2n}([\kk_1,\kk_m]).
\end{eqnarray}

\subsection{Extension of Reg PT: One-loop Level}

First, combining Equations~(\ref{power-r}) and (\ref{Reg1}),
the first-order contribution to the power spectrum in the $\Gamma$-expansion with Reg PT is given by
\begin{eqnarray}
		P_{\rm \Gamma}^{(1)}(z,k)=\left[ \Gamma^{(1)}(z,k) \right]^2P_{\rm L}(k) \to
		e^{-k^2 D^2\sigma_v^2}	\left[ 1 +  D^2 \Gamma_{\rm 1\mathchar`-loop}^{(1)}(k) + \frac{k^2 D^2 \sigma_v^2}{2}  \right]^2  D^2 P_{\rm L}(k).
		\label{WH1:1loop}
\end{eqnarray}

Next, we consider the second-order contribution calculated from Equation~(\ref{Reg1}):
\begin{eqnarray}
		P_{\rm \Gamma}^{(2)}(z,k)	&=& 2!\int \frac{d^3k_1}{(2\pi)^3}\int \frac{d^3k_2}{(2\pi)^3} (2\pi)^3 \delta_{D}(\kk-\kk_{[1,r]})
		\left[ \Gamma^{(2)}(z,\kk_1,\kk_2) \right]^2 P_{\rm L}(k_1) P_{\rm L}(k_2) \nonumber \\
		&\to&
		e^{-k^2 D^2\sigma_v^2}
	2! \int \frac{d^3k_1}{(2\pi)^3}\frac{d^3k_2}{(2\pi)^3}
	(2\pi)^3 \delta_D(\kk-\kk_{[1,2]}) D^2 P_{\rm L}(k_1)D^2 P_{\rm L}(k_2)\nonumber \\
	&& \times \Bigg[ \left( \Gamma_{\rm tree}^{(2)}(\kk_1,\kk_2) \right)^2
	+ 2 \Gamma_{\rm tree}^{(2)}(\kk_1,\kk_2)
	\left( D^2\Gamma_{\rm 1\mathchar`-loop}^{(2)}(\kk_1,\kk_2)
	+ \frac{k^2 D^2 \sigma_v^2}{2} \Gamma_{\rm tree}^{(2)}(\kk_1,\kk_2) \right)  \nonumber \\
	&& \hspace{5cm} + \left( D^2\Gamma_{\rm 1\mathchar`-loop}^{(2)}(\kk_1,\kk_2)
	+ \frac{k^2 D^2 \sigma_v^2}{2} \Gamma_{\rm tree}^{(2)}(\kk_1,\kk_2) \right)^2 \Bigg]. \nonumber \\
	\label{P_WH^2:1loop}
\end{eqnarray}
The first term in the square bracket leads to
\begin{equation}
		e^{-k^2D^2 \sigma_v^2} 2! \int \frac{d^3k_1}{(2\pi)^3}\frac{d^3k_2}{(2\pi)^3}
		(2\pi)^3 \delta_D(\kk-\kk_{[1,2]})
		\left[ F_2(\kk_1,\kk_2) \right]^2 D^2P_{\rm L}(k_1)D^2P_{\rm L}(k_2)= e^{-k^2D^2 \sigma_v^2}  D^4P_{22}(k).
		\label{WH2:1loop:part}
\end{equation}
If we only calculate Equations~(\ref{WH1:1loop}) and (\ref{WH2:1loop:part}),
we obtain a result called the one-loop Reg PT solution in \cite{Taruya:2012ut}:
\begin{eqnarray}
		P_{\rm Reg, 1\mathchar`-loop}(z,k) = e^{-k^2 D^2\sigma_v^2}
		\left[1 + D^2 \Gamma_{\rm 1\mathchar`-loop}^{(1)}(k) + \frac{k^2 D^2 \sigma_v^2}{2} \right]^2 D^2 P_{\rm L}(k)
		 + e^{-k^2D^2 \sigma_v^2}  D^4P_{22}(k).
		 \label{RegPT:1loop}
\end{eqnarray}
However, we further calculate the other terms in Equation~(\ref{P_WH^2:1loop}).
Although we need higher order terms in SPT than the one-loop level to strictly compute these terms, 
we here approximate them using $P_{13}\equiv 2P_{\rm L} \Gamma_{\rm 1\mathchar`-loop}^{(1)}$ and $P_{22}$, which are at the one-loop level.
The second and third terms in the square bracket in Equation~(\ref{P_WH^2:1loop}) are approximated for $k_1\to0$ as
\begin{eqnarray}
	&& e^{-k^2 D^2\sigma_v^2}
	 2! \int \frac{d^3k_1}{(2\pi)^3}\frac{d^3k_2}{(2\pi)^3}
	 (2\pi)^3 \delta_D(\kk-\kk_1-\kk_2)\big|_{k_1\to0} D^2P_{\rm L}(k_1) D^2 P_{\rm L}(k_2)\nonumber \\
	&& \times \Bigg[ 2 \Gamma_{\rm tree}^{(2)}(\kk_1,\kk_2)
	\left( D^2\Gamma_{\rm 1\mathchar`-loop}^{(2)}(\kk_1,\kk_2) 
	+ \frac{k^2 D^2 \sigma_v^2}{2} \Gamma_{\rm tree}^{(2)}(\kk_1,\kk_2) \right)\Bigg|_{k_1\to0}  \nonumber \\
	&& \hspace{2cm} + \left( D^2\Gamma_{\rm 1\mathchar`-loop}^{(2)}(\kk_1,\kk_2) 
	+ \frac{k^2 D^2 \sigma_v^2}{2} \Gamma_{\rm tree}^{(2)}(\kk_1,\kk_2) \right)^2 \Bigg|_{k_1\to0} \Bigg]\nonumber \\
	&=& e^{-k^2 D^2\sigma_v^2}
	 2! \int \frac{d^3k_1}{(2\pi)^3}\frac{d^3k_2}{(2\pi)^3}
	(2\pi)^3 \delta_D(\kk-\kk_2) D^2P_{\rm L}(k_1) D^2 P_{\rm L}(k_2)\nonumber \\
	&& \times \frac{1}{4}\left( \frac{\kk_1\cdot\kk_2}{k_1^2} \right)^2
	\Bigg[ 	2 \Gamma_{\rm tree}^{(1)}(k_2)
	\left( D^2\Gamma_{\rm 1\mathchar`-loop}^{(1)}(k_2) 
	+ \frac{k^2 D^2 \sigma_v^2}{2} \Gamma_{\rm tree}^{(1)}(k_2) \right)  \nonumber \\
	&& \hspace{5cm} + \left( D^2\Gamma_{\rm 1\mathchar`-loop}^{(1)}(k_2) 
	+ \frac{k^2 D^2 \sigma_v^2}{2} \Gamma_{\rm tree}^{(1)}(k_2) \right)^2 \Bigg]\nonumber \\
	&=&  e^{-k^2 D^2\sigma_v^2}\left( \frac{k^2 D^2 \sigma_v^2}{2} \right)
	\left[ \left(1 + D^2 \Gamma_{\rm 1\mathchar`-loop}^{(1)}(k) + \frac{k^2 D^2 \sigma_v^2}{2} \right)^2 D^2 P_{\rm L}(k)
	- D^2 P_{\rm L}(k)\right], 
	\label{PWH2:eq}
\end{eqnarray}
where we used the approximations for $\Gamma_{\rm 1\mathchar`-loop}^{(2)}$ and $\Gamma_{\rm tree}^{(2)}$ in Equation~(\ref{ap_G})
and $\delta_{\rm D}(\kk-\kk_{[1,2]})|_{k_1\to0} =\delta_{\rm D}(\kk-\kk_2)$.
Since we also need to consider the contribution from the region of $k_2 \to0$, we multiply Equation~(\ref{PWH2:eq}) by a factor of two.
Equation~(\ref{P_WH^2:1loop}) then becomes
\begin{eqnarray}
		P_{\rm \Gamma}^{(2)}(z,k)&\to& e^{-k^2D^2\sigma_v^2} D^4\left(P_{22}(k) - k^2 \sigma_v^2P_{\rm L}(k)  \right)  \nonumber \\
		&& + e^{-k^2D^2\sigma_v^2}(k^2 D^2 \sigma_v^2)
		\left(1 + D^2 \Gamma_{\rm 1\mathchar`-loop}^{(1)}(k) + \frac{k^2 D^2 \sigma_v^2}{2} \right)^2 D^2 P_{\rm L}(k).
	\label{WH2:1loop}
\end{eqnarray}

We apply the same analysis as above to an arbitrary order coefficient of the $\Gamma$-expansion.
The $r$th-order contribution to the power spectrum in the $\Gamma$-expansion is given from Equations~(\ref{power-r}) and (\ref{Reg1}) by
\begin{eqnarray}
	P_{\rm \Gamma}^{(r)}(z,k)&\to&e^{-k^2 D^2\sigma_v^2}
	r! \int \frac{d^3k_1}{(2\pi)^3} \cdots \frac{d^3k_r}{(2\pi)^3}
	(2\pi)^3 \delta_D(\kk-\kk_{[1,r]})D^2P_{\rm L}(k_1) \cdots D^r P_{\rm L}(k_r) \nonumber \\
	&& \times \Bigg[ \left[ \Gamma_{\rm tree}^{(r)}([\kk_1,\kk_r]) \right]^2
	+ 2 \Gamma_{\rm tree}^{(r)}([\kk_1,\kk_r])
	\left( D^{2}\Gamma_{\rm 1\mathchar`-loop}^{(r)}([\kk_1,\kk_r]) 
	+ \frac{k^2 D^2 \sigma_v^2}{2} \Gamma_{\rm tree}^{(r)}([\kk_1,\kk_r]) \right)  \nonumber \\
	&& \hspace{4cm} 
	+ \left( D^{2}\Gamma_{\rm 1\mathchar`-loop}^{(r)}([\kk_1,\kk_r]) 
	+ \frac{k^2 D^2 \sigma_v^2}{2} \Gamma_{\rm tree}^{(r)}([\kk_1,\kk_r]) \right)^2 \Bigg].
	\label{P_WH^r:1loop}
\end{eqnarray}
The first term in the square bracket is approximated as
\begin{eqnarray}
	&&\to e^{-k^2 D^2\sigma_v^2}
	r! \int \frac{d^3k_1}{(2\pi)^3} \cdots \frac{d^3k_r}{(2\pi)^3} (2\pi)^3 \delta_D(\kk-\kk_{[1,r]}) 
	D^2 P_{\rm L}(k_1) \cdots D^2P_{\rm L}(k_r) \nonumber \\
	&&\times \left[  \frac{r(r-1)}{2} \left[\Gamma_{\rm tree}^{(r)}([\kk_1,\kk_r]) \right]^2  \Big|_{k_3,\dots,k_r\to0} 
	+ \left( -r(r-1)  + r\right) \left[\Gamma_{\rm tree}^{(r)}([\kk_1,\kk_r]) \right]^2 \Big|_{k_2,\dots,k_r\to0}  \right]\nonumber \\
	&& = e^{-k^2D^2\sigma_v^2}\frac{(k^2D^2\sigma_v^2)^{r-2}}{(r-2)!} D^4\left(P_{22}(k) - k^2 \sigma_v^2 P_{\rm L}(k)  \right)
	 + e^{-k^2D^2\sigma_v^2}\frac{(k^2D^2\sigma_v^2)^{r-1}}{(r-1)!} D^2 P_{\rm L}(k),
\end{eqnarray}
where we multiply the first term by a factor of $r(r-1)/2$ because
we choose the two Fourier modes $\kk_1,\kk_2$ from the $r$ Fourier modes, $\kk_i$ $\{i=1,\cdots,r\}$.
We need the second term to exclude the redundant integrated regions defined by $k_2,\dots,k_r\to0$.
We used Equation~(\ref{ap_G}) 
and the approximation of the delta function: $\delta_{\rm D}(\kk-\kk_{[1,r]})|_{k_3,\dots,k_r\to0} = \delta_{\rm D}(\kk-\kk_{[1,2]})$
and $\delta_{\rm D}(\kk-\kk_{[1,r]})|_{k_2,\dots,k_r\to0} = \delta_{\rm D}(\kk-\kk_1)$.
The approximated solutions of the other terms in Equation~(\ref{P_WH^r:1loop}) 
are also calculated by the same derivation used in the case of $P_{\rm \Gamma}^{(2)}$:
\begin{eqnarray}
	e^{-k^2D^2\sigma_v^2}	\frac{(k^2D^2\sigma_v^2)^{r-1}}{(r-1)!}
	\left[\left( 1 + D^2 \Gamma_{\rm 1\mathchar`-loop}^{(1)}(k) + \frac{k^2 D^2 \sigma_v^2}{2} \right)^2 D^2P_{\rm L}(k) - D^2 P_{\rm L}(k)\right].
\end{eqnarray}
Therefore, we arrive at the approximated expression of $P_{\rm \Gamma}^{(r)}$:
\begin{eqnarray}
		P_{\rm \Gamma}^{(r)}(z,k)&\to &	
		  e^{-k^2D^2\sigma_v^2}
       \frac{(k^2D^2\sigma_v^2)^{r-1}}{(r-1)!}
	   \left( 1 + D^2 \Gamma_{\rm 1\mathchar`-loop}^{(1)}(k) + \frac{k^2 D^2 \sigma_v^2}{2} \right)^2 D^2 P_{\rm L}(k) \nonumber \\
        && +e^{-k^2D^2\sigma_v^2}
		\frac{(k^2D^2\sigma_v^2)^{r-2}}{(r-2)!} D^4\left(P_{22}(k) - k^2 \sigma_v^2P_{\rm L}(k)  \right) .
		\label{EXREG1}
		\end{eqnarray}
We propose this expression as an extended version of the one-loop Reg PT solution.
This certainly recovers the results in the cases of $r=1$ and $r=2$.

Note that if we sum up all orders of the $\Gamma$-expansion,
we derive the following power spectrum from the extended Reg PT at the one-loop level:
\begin{eqnarray}
		P(z,k) = \sum_{r=1}^{\infty} P_{\rm \Gamma}^{(r)}(z,k)
		= D^2 P_{\rm L}(k) + D^4P_{\rm 1\mathchar`-loop}(k) 
		+ \left( D^2 \Gamma_{\rm 1\mathchar`-loop}^{(1)}(k) + \frac{k^2 D^2 \sigma_v^2}{2}  \right)^2 D^2 P_{\rm L}(k),
		\label{Ex_SPT:1loop}
\end{eqnarray}
where $P_{\rm 1\mathchar`-loop}$ is the correction term at the one-loop level in SPT, defined as $P_{\rm 1\mathchar`-loop}\equiv P_{22} +P_{13}$.
\subsection{Extension of Reg PT: Two-loop Level}
Similar to the derivation used in Equation~(\ref{EXREG1}),
we derive the following approximated $P_{\rm \Gamma}^{(r)}$ from Equation~(\ref{Reg2}):
\begin{eqnarray}
	P_{\rm \Gamma}^{(r)}(z,k)&\to&e^{-k^2D^2\sigma_v^2}\frac{(k^2D^2\sigma_v^2)^{r-1}}{(r-1)!}
	\Bigg[ 1 + \left( D^2 \Gamma_{\rm 1\mathchar`-loop}^{(1)}(k) + \frac{k^2D^2\sigma_v^2}{2} \right)  \nonumber \\
	&& \hspace{3cm}+\left( D^4\Gamma_{\rm 2\mathchar`-loop}^{(1)}(k) + \frac{k^2D^2\sigma_v^2}{2} D^2\Gamma_{\rm 1\mathchar`-loop}^{(1)}(k)
	+ \frac{1}{2}\left( \frac{k^2D^2\sigma_v^2}{2} \right)^2	\right) \Bigg]^2D^2P_{\rm L}(k) \nonumber \\
		&& + e^{-k^2D^2\sigma_v^2}\frac{(k^2D^2\sigma_v^2)^{r-2}}{(r-2)!}
		D^4\left[ P_{22}(k) - k^2 \sigma_v^2 P_{\rm L}(k)\right] \nonumber \\
		&& + e^{-k^2D^2\sigma_v^2} \frac{(k^2D^2\sigma_v^2)^{r-2}}{(r-2)!}
		D^6\left[ P_{24}(k) + k^2 \sigma_v^2 P_{22}(k) - k^2 \sigma_v^2 \left( P_{13}(k) + k^2 \sigma_v^2 P_{\rm L}(k) \right) \right] \nonumber \\
		&&+  e^{-k^2D^2\sigma_v^2}\frac{(k^2D^2\sigma_v^2)^{r-2}}{(r-2)!}
          D^8\Bigg[ P_{44a}(k) + \frac{k^2 \sigma_v^2}{2} P_{24}(k) + \frac{(k^2 \sigma_v^2)^2}{4} P_{22}(k) \nonumber \\
		  &&\hspace{5cm}
		  - k^2 \sigma_v^2 \left( \Gamma_{\rm 1\mathchar`-loop}^{(1)}(k) + \frac{k^2  \sigma_v^2}{2}  \right)^2 P_{\rm L}(k)\Bigg] \nonumber \\
		&& + e^{-k^2D^2\sigma_v^2}\frac{(k^2D^2\sigma_v^2)^{r-3}}{(r-3)!}
		D^6\left[ P_{33b}(k) - k^2 \sigma_v^2 P_{22}(k) + \frac{(k^2 \sigma_v^2)^2}{2} P_{\rm L}(k) \right],
		\label{ExtendedRegPT:2loop}
\end{eqnarray}
where $P_{24}$, $P_{33b}$, and $P_{44a}$ are defined as
\begin{eqnarray}
		P_{24}(k)  &\equiv&  24	\int \frac{d^3k_1}{(2\pi)^3}\frac{d^3k_2}{(2\pi)^3}\frac{d^3p}{(2\pi)^3} 
		(2\pi)^3 \delta_{D}(\kk-\kk_{[1,2]}) F_2(\kk_1,\kk_2) F_4(\kk_1,\kk_2,\pp,-\pp)P_L(p) P_{\rm L}(k_1) P_{\rm L}(k_2),\nonumber \\
		P_{33b}(k) &\equiv& 6	\int \frac{d^3k_1}{(2\pi)^3}\frac{d^3k_2}{(2\pi)^3}\frac{d^3k_3}{(2\pi)^3} 
		(2\pi)^3 \delta_D(\kk-\kk_{[1,3]}) \left[  F_3(\kk_1,\kk_2,\kk_3)\right]^2 P_{\rm L}(k_1) P_{\rm L}(k_2) P_{\rm L}(k_3), \nonumber \\
		P_{44a}(k) &\equiv&  72 \int \frac{d^3k_1}{(2\pi)^3}\frac{d^3k_2}{(2\pi)^3} \frac{d^3p_1}{(2\pi)^3}\frac{d^3p_2}{(2\pi)^3}
		(2\pi)^3 \delta_D(\kk-\kk_{[1,2]}) \nonumber \\
		&& \times F_4(\kk_1,\kk_2,\pp_1,-\pp_1)F_4(\kk_1,\kk_2,\pp_2,-\pp_2) P_{\rm L}(k_1)P_{\rm L}(k_2) P_{\rm L}(p_1) P_{\rm L}(p_2).
		\label{corrections:2loop}
\end{eqnarray}

Equation~(\ref{ExtendedRegPT:2loop}) recovers the two-loop Reg PT solution which is given by \citep{Taruya:2012ut}
\begin{eqnarray}
		 P_{\rm Reg,2\mathchar`-loop}(z,k) 
		&=&  e^{-k^2D^2\sigma_v^2}
        \Bigg[ 1 + \left( D^2 \Gamma_{\rm 1\mathchar`-loop}^{(1)}(k) + \frac{k^2D^2\sigma_v^2}{2} \right)  \nonumber \\
		&&\hspace{2cm} +\left( D^4\Gamma_{\rm 2\mathchar`-loop}^{(1)}(k) + \frac{k^2D^2\sigma_v^2}{2} 
		D^2\Gamma_{\rm 1\mathchar`-loop}^{(1)}(k)
	       + \frac{1}{2}\left( \frac{k^2D^2\sigma_v^2}{2} \right)^2	\right) \Bigg]^2D^2P_{\rm L}(k) \nonumber \\
		&&+ e^{-k^2D^2\sigma_v^2}
		\Bigg[  D^4P_{22}(k) + D^6\left( P_{24}(k) + k^2 \sigma_v^2 P_{22}(k)\right)  \nonumber \\
		&&\hspace{3cm}
		+ D^8\left( P_{44a}(k) + \frac{k^2 \sigma_v^2}{2} P_{24}(k) + \frac{(k^2 \sigma_v^2)^2}{4}P_{22}(k) \right)\Bigg]  \nonumber \\
		&&+ e^{-k^2D^2\sigma_v^2} P_{33b}(k).
		\label{RegPT:2loop}
\end{eqnarray}
In the case of $r=1$, Equation~(\ref{ExtendedRegPT:2loop}) coincides with the first term in Equation~(\ref{RegPT:2loop}).
For $r=2$ and $r=3$, we have additional terms compared to the second and third terms in Equation~(\ref{RegPT:2loop}).
This is because 
Equation~(\ref{ExtendedRegPT:2loop}) is computed using Reg PT at the two-loop level (Equation~(\ref{Reg2})),
although the second and third term in Equation~(\ref{RegPT:2loop}) are calculated
by Reg PT at the one-loop level (Equation~(\ref{Reg1})) and the tree level, respectively.
Furthermore, we can also derive higher order terms than the third order in the $\Gamma$-expansion from Equation~(\ref{ExtendedRegPT:2loop}).

The power spectrum including all orders of the $\Gamma$-expansion is given by
\begin{eqnarray}
		P(z,k) &=&  D^2 P_L(k) + D^4P_{\rm 1\mathchar`-loop}(k) + D^6 P_{\rm 2\mathchar`-loop}(k) \nonumber \\
		&&	+ 2D^8\left( \Gamma_{\rm 1\mathchar`-loop}^{(1)}(k) + \frac{k^2\sigma_v^2}{2}\right)
		\left( \Gamma_{\rm 2\mathchar`-loop}^{(1)}(k) + \frac{k^2\sigma_v^2}{2} \Gamma_{\rm 1\mathchar`-loop}^{(1)}(k)
		+ \frac{1}{2}\left( \frac{k^2\sigma_v^2}{2} \right)^2\right) P_{\rm L}(k) \nonumber \\
    &&+ D^8\left[ P_{44a}(k) + \frac{k^2 \sigma_v^2}{2} P_{24}(k)+ \frac{(k^2 \sigma_v^2)^2}{4} P_{22}(k)  
	- k^2 \sigma_v^2 \left( \Gamma_{\rm 1\mathchar`-loop}^{(1)}(k) + \frac{k^2  \sigma_v^2}{2}  \right)^2D^2P_{\rm L}(k)\right] \nonumber \\
	&& + D^{10}\left( \Gamma_{\rm 2\mathchar`-loop}^{(1)}(k) + \frac{k^2\sigma_v^2}{2} \Gamma_{\rm 1\mathchar`-loop}^{(1)} (k)
	+ \frac{1}{2}\left( \frac{k^2\sigma_v^2}{2} \right)^2\right)^2  P_{\rm L}(k), 
	\label{SPT2loop+}
\end{eqnarray}
where $P_{\rm 2\mathchar`-loop}$ is the correction term at the two-loop level in SPT, defined as
\begin{eqnarray}
		P_{\rm 2\mathchar`-loop} \equiv P_{15}(k) + P_{24}(k) + P_{33a}(k) + P_{33b}(k),
\end{eqnarray}
with $P_{15}\equiv 2\Gamma_{\rm 2\mathchar`-loop}^{(1)} P_{\rm L}$ and $P_{33a}\equiv [\Gamma_{\rm 1\mathchar`-loop}^{(1)}]^2 P_{\rm L}$.
As in the case of the extended 1-loop Reg PT, 
we obtain the two-loop SPT solution with the additional correction terms.

\section{Modified SPT}
\label{Ex_SPT}

\begin{figure}[t]
				\begin{center}
						\epsscale{0.7}
						\plotone{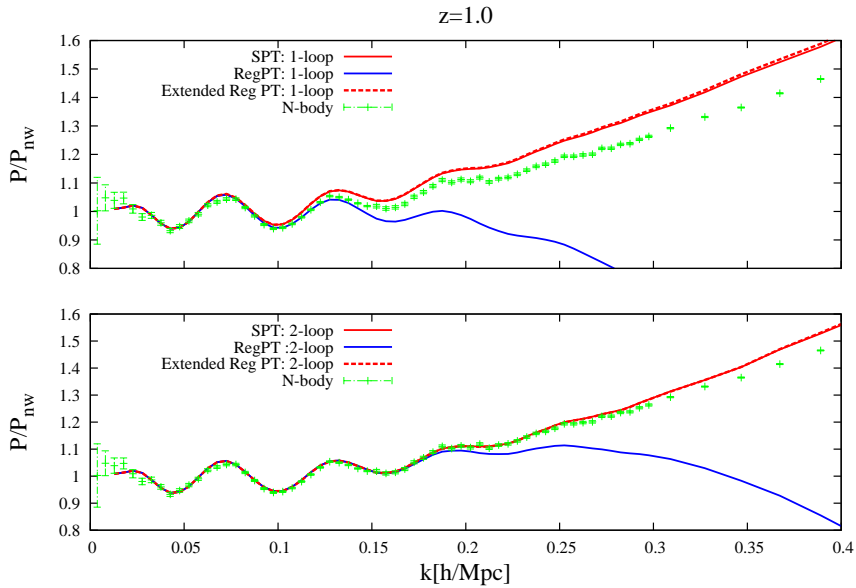}
				\end{center}
		\caption{Ratios of the predicted power spectra $P$ to the smoothed linear power spectrum without BAO $P_{\rm nw}$
		presented by \cite{Eisenstein:1997ik} are plotted at $z=1.0$.
		The predicted power spectra are computed by 
		SPT (red solid), Reg PT (blue solid, Equations~(\ref{RegPT:1loop}) and (\ref{RegPT:2loop})), 
		the extended Reg PT (red dashed, Equations~(\ref{Ex_SPT:1loop}) and (\ref{SPT2loop+})), and $N$-body simulations (green symbols).
		}
		\label{fig:SG}
\end{figure}

We investigate the relation between the results from Reg PT (Equations~(\ref{RegPT:1loop}) and~(\ref{RegPT:2loop})),
the extended Reg PT (Equations~(\ref{Ex_SPT:1loop}) and~(\ref{SPT2loop+})), and SPT.
One might think that the results of the extended RegPT
give a better prediction for the nonlinear matter power spectra than those of SPT,
because the extended RegPT has the correction terms as well as the SPT solution.
However, this is not true.
Figure~\ref{fig:SG} shows that the additional terms in Equations~(\ref{Ex_SPT:1loop}) and~(\ref{SPT2loop+}) 
hardly contribute to the predicted nonlinear power spectra (see the red solid and dashed lines in Figure~\ref{fig:SG}).
Therefore, we do not need to consider the partial correction terms at higher order than the two-loop level in SPT, 
such as $P_{44a}$, $\Gamma_{\rm 1\mathchar`-loop}^{(1)}\Gamma_{\rm 2\mathchar`-loop}^{(1)}$, and 
$\left[ \Gamma_{\rm 2\mathchar`-loop}^{(1)} \right]^2$ in  Equation~(\ref{SPT2loop+}).
Moreover, we clearly find that the predictions from Reg PT (blue lines)
lose information on the nonlinear evolution of dark matter compared to the SPT predictions (red lines)
due to the truncation of the $\Gamma$-expansion at second or third order.

\begin{figure}[t]
		\epsscale{0.80}
		\plotone{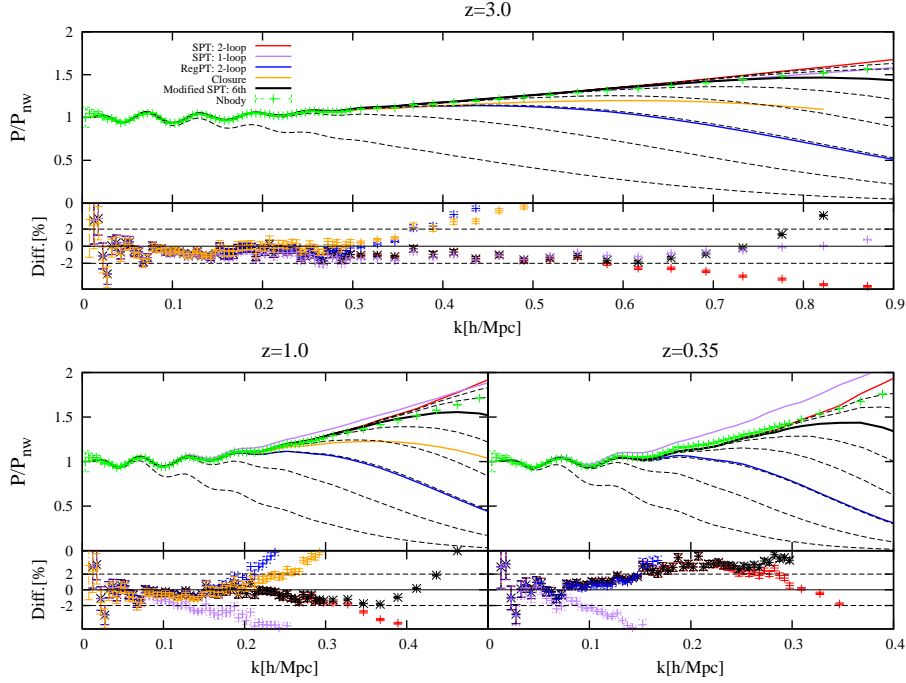}
		\caption{Ratios of the predicted power spectra $P$ to the smoothed linear power spectrum $P_{\rm nw}$
		are plotted at the redshifts of $z=3.0$ 1.0, and 0.35.
		The predicted power spectra are computed by 
		the one-loop SPT (purple solid), the two-loop SPT (red solid), the two-loop Reg PT in Equation~(\ref{RegPT:2loop}) (blue solid),
		the closure theory (orange solid), the $N$-body simulations (green symbols),
		and the modified two-loop SPT (Equation~(\ref{ExSPT:2loop}))
		where the finite truncated orders of the $\Gamma$-expansion go from 1 to 8
		(black dashed: $r_k=1\mathchar`-5$, 7, and 8 from left to right in order; black solid: $r_k=6$).
		Furthermore, the fractional differences, defined as ${\rm Diff[\%]} \equiv (P_{\rm Nbody}-P)*100/P_{\rm Nbody}$
		where $P_{\rm Nbody}$ is the predicted power spectrum by the $N$-body simulations,
		are plotted at each redshift.
		There, the plotted models are the one-loop SPT (purple symbols), 
		the two-loop SPT (red symbols), the two-loop Reg PT (blue symbols), the closure theory (orange symbols),
		and the modified SPT truncated at the sixth order of the $\Gamma$-expansion (black symbols).
		}
		\label{fig:power}
\end{figure}

Now, we present the following modified SPT solution at the one- and two-loop level:
\begin{eqnarray}
		P_{\rm SPT,1\mathchar`-loop}(z,k) &=&	 e^{-k^2D^2\sigma_v^2}
        \sum_{r=1}^{r_k} \frac{(k^2D^2\sigma_v^2)^{r-1}}{(r-1)!}
	   \left( D^2 P_{\rm L}(k) + D^{4}\left( P_{13}(k) + k^2 \sigma_v^2 P_{\rm L}(k) \right) \right) \nonumber \\
	   && + e^{-k^2D^2\sigma_v^2}
		\sum_{r=2}^{r_k}\frac{(k^2D^2\sigma_v^2)^{r-2}}{(r-2)!} D^4\left(P_{22}(k) - k^2 \sigma_v^2P_{\rm L}(k)  \right) 
		\label{ExSPT:1loop}
\end{eqnarray}
\begin{eqnarray}
		&&P_{\rm SPT,2\mathchar`-loop}(z,k) \nonumber \\
		&&=e^{-k^2D^2\sigma_v^2}  \sum_{r=1}^{r_k}\frac{(k^2D^2\sigma_v^2)^{r-1}}{(r-1)!}
		\Bigg[ D^2 P_{\rm L}(k) + D^4\left( P_{13}(k) + k^2 \sigma_v^2 P_L(k)  \right) \nonumber \\
		&&\hspace{5cm} + D^6\left( P_{33a}(k) + P_{15}(k) +	k^2 \sigma_v^2 P_{13}(k) + \frac{(k^2 \sigma_v^2)^2}{2} P_{\rm L}(k)\right) \Bigg] 
		\nonumber \\
		&& + e^{-k^2D^2\sigma_v^2}  \sum_{r=2}^{r_k} \frac{(k^2D^2\sigma_v^2)^{r-2}}{(r-2)!}
		\Bigg[D^4\left( P_{22}(k) - k^2 \sigma_v^2 P_{\rm L}(k)\right) \nonumber \\
		&&\hspace{5cm} 
		D^6\left( P_{24}(k) + k^2 \sigma_v^2 P_{22} - k^2 \sigma_v^2 \left( P_{13}(k) + k^2 \sigma_v^2 P_{\rm L}(k) \right)  \right) \Bigg]
		\nonumber \\
		&& + e^{-k^2D^2\sigma_v^2}  \sum_{r=3}^{r_k}\frac{(k^2D^2\sigma_v^2)^{r-3}}{(r-3)!}
		D^6\left[ P_{33b}(k) - k^2 \sigma_v^2 P_{22}(k) + \frac{(k^2 \sigma_v^2)^2}{2} P_{\rm L}(k) \right].
		\label{ExSPT:2loop}
\end{eqnarray}
These expressions are obtained by removing the additional terms at higher orders than the one- and two-loop level in SPT
from Equations~(\ref{EXREG1}) and (\ref{ExtendedRegPT:2loop}), respectively, and 
cutting off the $\Gamma$-expansion at a finite order $r_k$.
For $r_k\to\infty$, the above solutions reduce to the usual SPT solutions.
Note that these solutions avoid the divergent behavior of the power spectrum at small scales
by considering the finite order $r_k$ one needs.

In Figure~\ref{fig:power}, 
we show a comparison between the predicted power spectra from 
the modified SPT solution at the two-loop level (Equation~(\ref{ExSPT:2loop}))
with the finite truncated orders of the $\Gamma$-expansion going from $1$ to $8$ (black dashed),
the one-loop SPT (purple solid), the two-loop SPT (red solid),
the two-loop Reg PT (blue solid), the closure theory \citep{Taruya:2009ir} (orange solid), and the $N$-body simulations (green symbols).
First, the predictions of the two-loop SPT agree better with the $N$-body simulations 
than those of the two-loop Reg PT and the closure theory at the redshifts ($z=3.0$, 1.0, and 0.35).
Next, the modified SPT solutions truncated at the third order of the $\Gamma$-expansion give almost the same results as the two-loop Reg PT.
As previously mentioned, this is because the additional correction terms in the two-loop Reg PT 
have no effective information on the nonlinear evolution of dark matter.
Third, the divergent behavior in the two-loop SPT at small scales is indeed modified by the truncation of the $\Gamma$-expansion
(see the black dashed lines).
We emphasized the sixth-order solutions of the $\Gamma$-expansion in the modified SPT using the black solid lines.
The sixth-order solutions accidentally agree with the $N$-body results better than the two-loop SPT solutions
due to the damping factor from the truncated $\Gamma$-expansion 
(see the red and black symbols in the fractional differences in Figure~\ref{fig:power}).
We suggest that this solution should be used instead of the two-loop SPT.
The solution guarantees accuracy comparable to the two-loop SPT and has no divergent behavior at small scales.

\section{Conclusion}
\label{Conclusion}

Based on the idea that any resummation theory should be described and solved through the kernel function in SPT,
we established an approximation of the kernel function (Equation~(\ref{ap_FG})).

This approximation explains the existing Reg PT model, and gives its natural extension (Equation~(\ref{ExtendedRegPT:2loop})).
However, the extended Reg PT is equivalent to SPT with negligibly small correction terms,
and gives almost the same result as for the SPT (see Figure~\ref{fig:SG}).
Therefore, we do not need to consider these additional terms.
Moreover, since the Reg PT solutions (Equations~(\ref{RegPT:1loop}) and (\ref{RegPT:2loop}))
truncate the $\Gamma$-expansion at the second and third order,
they do not contain more effective information on the nonlinear dark matter evolution than the SPT solutions.
In other words, while the original Reg PT solution becomes zero beyond BAO scales due to exponential damping,
the extended Reg PT solution allows us to predict the power spectrum at the high-$k$ region,
because the exponential damping factor is completely canceled out.
In fact, the nonlinear power spectra computed in the two-loop SPT agree better with the $N$-body results 
than that of the two-loop Reg PT at redshifts of $z=3.0$, 1.0, and 0.35, especially at high-$k$ region.
Hence, the two-loop SPT is one of the best models to predict the nonlinear matter power spectrum 
even at present (see Figure~\ref{fig:power}).

In the spirit of the $\Gamma$-expansion,
each loop corrections are concentrated at a narrow range of scales, and the reliable range of the approximation becomes much more obvious.
Therefore, the Reg PT two-loop solution, which truncates the $\Gamma$-expansion at the third order,
guarantees the validity of the predicted power spectrum at small scale regions.
In this paper, we showed that the Reg PT one- and two-loop solutions are a part of the SPT one- and two-loop solutions plus negligible terms
(Equation~(\ref{SPT2loop+}))
and extract the information at small scale regions in the SPT solutions.
Thereby, the predicted power spectra in the Reg PT model have the same behavior as those in the SPT model at the BAO scales
(see Figure~\ref{fig:power}).
Furthermore, we interpret the truncation of the $\Gamma$-expansion as the loss of the information at small scale regions.
In this sense the Reg PT solution does not have more effective information on the nonlinear power spectrum than the SPT solution.

We presented modified versions of the SPT solutions at
the one- and two-loop level (Equations~(\ref{ExSPT:1loop}) and (\ref{ExSPT:2loop})),
which are the SPT solutions described in the context of the $\Gamma$-expansion.
These solutions avoid the divergent behavior of the SPT solutions at small scales
by the truncation of the $\Gamma$-expansion at the finite order one needs.
We suggest that one should use the modified SPT solution at the sixth order of the $\Gamma$-expansion instead of the two-loop SPT solution,
because this solution guarantees accuracy comparable to the two-loop SPT and has no divergent behavior.

Finally, we shall mention some applications of our result.
So far the correlation function has not been computed from SPT because of the divergent behavior.
However, the modified SPT solution would allow us to do it.
Similarly, we occasionally 
encounter the difficulty that we have to compute the integral of the linear or nonlinear power spectrum but the integrand does not converge,
such as the calculation of back reaction from the matter perturbations.
Our result would be useful for this case.
However, we leave these applications as our future work.

\acknowledgments
We thank T. Nishimichi and A. Taruya for providing numerical simulation results and useful comments.
We also thank L. Mercoli for checking this paper.
This work is supported in part by the GCOE Program  ``Weaving Science Web beyond Particle-Matter Hierarchy''
at Tohoku University and by a Grant-in-Aid for Scientific Research from JSPS (No. 24-3849 for N.S.S. and No. 20540245 for T.F.).


\end{document}